# Accurately Quantifying Radiative Cooling Potentials: A Temperature-correction to the Transmittance-based approximation

Jyotirmoy Mandal,*[†] Xin Huang,[†] Aaswath P. Raman*
Department of Materials Science, University of California, Los Angeles
Correspondence: jyotirmoymandal@g.ucla.edu, aaswath@ucla.edu
[†] These authors contributed equally to the work.

## Abstract

Theoretical calculations of the cooling potential of radiative cooling materials are crucial for determining their cooling capability under different meteorological conditions and evaluating their performance. To enable these calculations, accurate models of long-wave infrared downwelling atmospheric irradiance are needed, However, the transmittance-based cosine approximation, which is widely used to determine radiative cooling potentials, does not account for the cooling potential arising from heat loss to the colder reaches of the atmosphere itself. Here, we show that use of the approximation can lead to > 10% underestimation of the cooling potential relative to MODTRAN 6 outputs. We propose a temperature correction to the transmittance-based approximation which accounts for heat loss to the cold upper atmosphere, and significantly reduces this underestimation, while retaining the advantages of the original model. In light of the widespread and continued use of the transmittance-based model, our results highlight an important source of potential errors and a means to correct for them.

## Introduction

In recent years, radiative cooling has seen growing scientific and commercial interest for applications ranging from the passive cooling of buildings to geoengineering. The process, which involves a spontaneous heat loss from terrestrial objects to the atmosphere and outer space by radiation of heat (and reflection of incident sunlight) through the atmospheric transmission windows, has a zero-energy, zero-carbon functionality and a net cooling effect on the environment.[1] Precisely how much cooling occurs for a given surface depends strongly on meteorological conditions. For instance, the cooling potential, which is defined as the difference between the radiance from a sky-facing radiative cooler and the downwelling atmospheric irradiance, can vary between ~0-150 $Wm^{-2}$ depending on the ambient temperature and total precipitable water (TPW) content.[2]

Given this large variability, accurate determination of radiative cooling potentials is crucial for validating the performance of radiative coolers, informing industry on the geographical scope of designs like cool-roof paints,[3] and potentially the best geographic regions for radiative cooling approaches for geoengineering.[2,4] However, standard models used today to calculate radiative cooling potentials or cooling power often rely on simplified models of atmospheric response that are used beyond their scope, leading to systematic errors. The most prevalent example of this is the transmittance-based cosine approximation, widely used to model radiative cooling potentials, particularly of materials with strong spectral and angular selectivity in their emissivity.

In this paper, we elucidate the source of the errors in atmospheric radiance calculations that use the transmittance-based cosine approximation, and demonstrate its underestimation of radiative cooling potential due to its simplified accounting of the irradiance from greenhouse gases where the atmosphere is transparent. Comparative analysis against the MODTRAN atmospheric hemispherical irradiance model shows that the transmittance-based cosine approximation yields a significantly higher downwelling atmospheric irradiance and thus cooling potentials that are lower by 6-24 Wm$^{-2}$ under typical operating conditions, which is 10-23% more than the approximation itself. To address this, we apply a temperature correction that accounts for the high elevations and thus low temperatures of greenhouse gases, namely water vapor, carbon dioxide and ozone, which allows a net heat transfer to them from the earth's surface. This reduces the underestimation of the cooling potential to 0.1-6%, while retaining the useful angular resolution of the transmittance-based cosine approximation, which the MODTRAN hemispherical irradiance model does not provide. Our results suggest that recently constructed maps of radiative cooling potentials may require corrections. Moreover, they indicate that the common use of the uncorrected transmittance-based cosine approximation to verify experimental demonstrations of radiative cooling could be leading to an overestimation of performance of radiative cooling designs across the literature.

## Atmospheric Irradiance and the Transmittance-based Cosine Approximation

Due to its constituent greenhouse gases that are intrinsically absorptive or emissive in the thermal wavelengths, the atmosphere radiates heat towards the earth's surface. The difference between this irradiance $I_{atm}$ and the blackbody radiance $I_{BB}$ ($T_{amb}$) at the ambient air temperature ($T_{amb}$) close to the ground is what is typically defined as the cooling potential or cooling power $P_{cooling}$ The cooling potential arises primarily within the long-wavelength infrared (LWIR, 8-13 $\mu m$) atmospheric transmission window, where the low intrinsic absorption of water vapour lowers $I_{atm}$, and reveals the cold upper reaches of the atmosphere and the cold space beyond, allowing for heat radiated upwards from the ground to be lost.

While the calculation of the blackbody irradiance is straightforward, it is more challenging to calculate $I_{atm}$, which needs to account for, at the very least, the spectral properties of greenhouse gases, their distributions along the height of the atmosphere, and variations in temperature across heights. A long history of work on this topic has yielded a number of useful theoretical models for calculating $I_{atm}$, each of which is reasonably accurate within its scope of use.[1,5–9] For instance, when using a spectrally flat emitter (gray- or black-body), simple correlations have been shown to be very accurate in predicting downwelling atmospheric irradiance.[7] However, for many radiative cooling calculations, spectrally and angularly resolved sky irradiance is often required since radiative cooling surfaces can present highly spectrally and angularly selective emissivity . Models which yield the level of detail needed for such calculations are comparatively rare.[1,5,10] One model, which has achieved almost universal use in recent radiative cooling literature is the transmittance-based cosine approximation,[1,11–26] which was first used as part of a more comprehensive model by Granqvist in 1981.[1] This model assumes that the irradiance of the atmosphere originates from the greenhouse gases water vapor, carbon dioxide and ozone, and calculates the spectral, angular sky irradiance based on an effective spectral angular emittance as follows:

$$I_{atm}(\theta, \lambda, T_{amb}) = \varepsilon_{atm}(\theta, \lambda, T_{amb}) \cdot I_{BB}(\theta, \lambda, T_{amb}) \quad (1)$$

where

$$\varepsilon_{atm}(\theta, \lambda, T_{amb}) = 1 - [1 - \varepsilon_{atm}(0, \lambda, T_{amb})]^{1/\cos\theta} \quad (2)$$

Here, $\varepsilon_{atm}$ is the effective emittance of the atmosphere, $\theta$ is the angle measured from the zenith and $\lambda$ is the wavelength. As stated by Granqvist, implicit in this model is Kirchhoff's Law, which states that at thermal equilibrium:

$$\varepsilon_{atm}(\theta, \lambda) = \alpha_{atm}(\theta, \lambda) = 1 - \tau_{atm}(\theta, \lambda) \quad (3)$$

Where $\alpha_{atm}(\theta, \lambda)$ is the spectral, directional absorptance of the atmosphere. $\tau_{atm}(\theta, \lambda)$ is the spectral, directional transmittance, and is calculated based on the zenith-ward value $\tau_{atm}(0, \lambda)$ using the cosine approximation:

$$\tau_{atm}(\theta, \lambda) = \tau_{atm}(0, \lambda)^{\frac{1}{\cos\theta}} \quad (4)$$

The hemispherical irradiance $I_{atm}$ can be calculated by hemispherical integration of the angular values from (1), and in turn yields $P_{cooling} = I_{BB} - I_{atm}$.

### Issues with the Transmittance-based Cosine Approximation

While the model provides a reasonable estimate of $I_{atm}(\theta, \lambda)$, a crucial point to note is that in using Kirchhoff's law and equation (3), it implicitly assumes that the atmosphere is a homogenous entity at thermal equilibrium.[1] In reality, it is a complex, semi-transparent structure whose temperature decreases with height. Where the atmosphere is particularly transparent, the colder upper reaches of the atmosphere are apparent from the ground, which enables heat loss not only through the atmosphere, but also to it (Fig. 1A), a fact that is not accounted for by the transmittance-based approximation.

The case for ozone, illustrates this well. It is first important to note that the overwhelming majority of ozone occurs in frigid 10-40 km heights of the stratosphere, in what is known as the ozone layer. Consequently, any downwelling irradiance outside the LWIR window from ozone is masked by the highly absorptive water vapor and $CO_2$ in the troposphere. In the LWIR window, however, the transparency of the atmosphere reveals the ozone layer and outer space beyond. This means that some of the radiance from the earth's surface is absorbed by the ozone, and much of its intrinsic radiance at ~ 9.5 μm reaches the earth. However, because of the low temperature of the ozone layer, which can be ~70°C lower than $T_{amb}$ at altitudes where ozone concentration peaks in the winter, and ~40°C lower in the winter,[28] it radiates far less towards the ground than it absorbs from the latter. In other words, a net heat loss occurs from the surface to the ozone layer, and the assumption of a thermally homogeneous atmosphere no longer holds (Fig. 1A). This is not captured by Equation (3), which implicitly assumes the ozone layer to be at $T_{amb}$, leading the downwelling irradiance from ozone to be incorrectly equal to the fraction of $I_{BB}(T_{amb})$ absorbed by it. Consequently, heat loss to the ozone layer is not registered, $I_{atm}$ is overestimated, and $P_{cooling}$ is underestimated, as is shown in Fig. 1B. This is also evident when one compares the hemispherical emittance calculated by MODTRAN with $I_{atm}$ calculated using the transmittance-based cosine approximation. The difference due to the ozone effect alone is about 5-18.5 Wm$^{-2}$ depending on the atmosphere type and temperature, which is a significant 6-21% of the net cooling potential predicted by the transmittance-based cosine approximation.

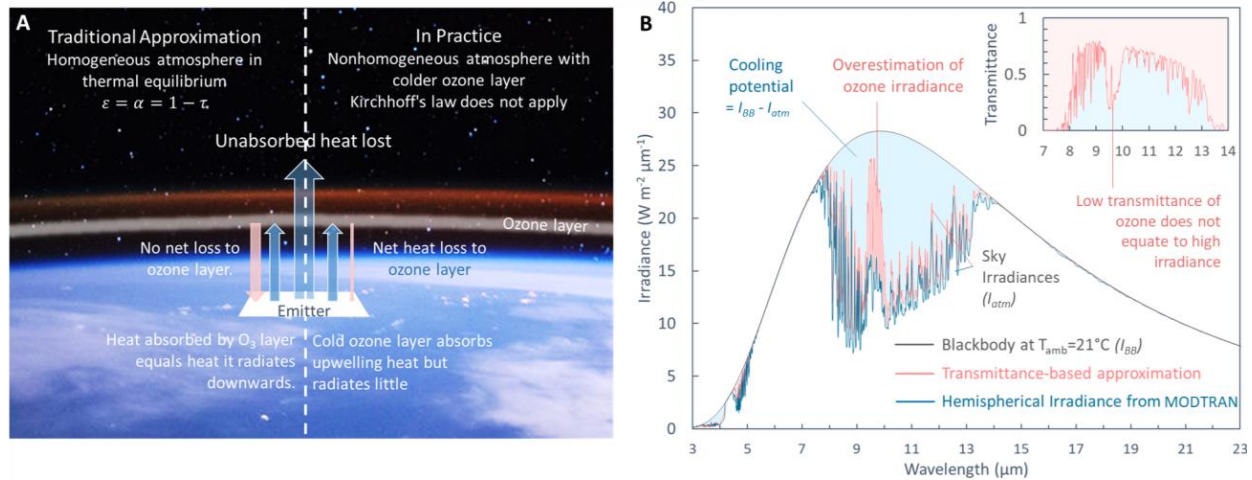

**Figure 1.** A) Conceptual framework for traditional atmospheric thermal radiation against real case atmospheric thermal radiation. The photograph[27] was used under the Creative Commons-CC-BY-NC-ND 2.0 License. B) The spectral hemispherical atmospheric irradiance of the transmittance-based approximation and that from MODTRAN,

A similar effect occurs for water vapor and carbon dioxide, which we consider collectively in this analysis. The two gases are well mixed throughout the atmosphere, and a majority of their downwelling irradiance arises from within ~10$^2$ m depths of the atmosphere near the earth's surface outside the LWIR window (which is at ~$T_{amb}$), and within 2 km of the earth's surface (or within ~ 12 °C of $T_{amb}$) in the LWIR. The resulting temperature differences are far less than that for ozone. Consequently, the transmittance-based approximation is largely correct outside the LWIR window where the atmosphere is opaque, and shows a lower underestimation of $P_{cooling}$ than seen for ozone at individual LWIR wavelengths (Fig. 1B). However, unlike for ozone, the underestimation occurs over a much broader bandwidth – across the LWIR and for dry atmospheres, across the 16-20 μm wavelengths, which adds to a significant total when integrated.

It should be noted that Granqvist explicitly proposed the transmittance-based cosine approximation for use with a box model for calculating the radiative cooling powers of SiO films on metal. Since the irradiance from ozone and absorptance/emittance of SiO films have little overlap, and the SiO film has a narrowband emittance, such a choice is justifiable in that context. However, the approximation has since been used to calculate radiative cooling potentials of ideal emitters and cooling powers of radiative coolers with different spectral emittances, leading both to a systematic underestimation of the cooling potential, and relative to it, an overestimation of the performance. [11–26] The MODTRAN hemispherical emittance, which is more accurate, should ideally be used instead. Indeed, Granqvist used a similar model in a later work.[5] However, because publicly available versions of MODTRAN contain no angle resolved information,[28,29] to date, it has mostly been used for ideal emitters or real ones with ultra-wide angle emittances.[30] Furthermore, given the widespread use of the transmittance-based model in the radiative cooling community, it may be expedient for researchers to use a modified version of the model that provides the flexibility needed to accurately model spectrally and angularly-selective radiative coolers. Towards that end, in the subsequent section, we propose a correction to the transmittance-based cosine approximation that reduces the systematic overestimation of the ozone's irradiance, while retaining the necessarily angular resolution of the model.

## Temperature-Corrections of the Transmittance-based Model

Although the publicly-accessible MODTRAN hemispherical irradiance model provides a highly accurate estimate of the cooling potential, it does not contain angle-resolved information, which is crucial for calculating the performance of typical radiative coolers whose emittances can vary considerably with angle,[1,12,31,32] and which the transmittance-based cosine approximation provides. Therefore, here we propose a correction that retains the mathematical fundamentals of the original model, but corrects for the overestimation of the irradiance from greenhouse gases. To do so, the lower effective temperatures of the ozone layer, and water vapor+$CO_2$ in the LWIR, which allow for heat loss from the terrestrial environment, must be taken into account. The ozone layer is kilometers thick, and $CO_2$ and water vapor are distributed throughout the atmosphere, which means that their radiative contributions are determined by a temperature distribution along their height. However, we can simplify calculations assuming that the irradiance of the i[th] gaseous component arises from a specific combination of its emittance $\varepsilon_i$ and effective temperature $T_i$. The irradiance $I_{atm}$ can then be separated into two contributions, one for ozone, which is distributed high in the atmosphere, and one for $CO_2$ +water vapor, which is distributed throughout, as follows:

$$I_{atm}(\theta,\lambda) = \varepsilon_{ozone}(\theta,\lambda) \times I_{BB}(T_{ozone},\lambda) + \varepsilon_{rest}(\theta,\lambda) \times I_{BB}(T_{rest},\lambda) \qquad (5)$$

The directional emittances $\varepsilon_i(\theta,\lambda)$ of each component is calculated using Equations 2-4, using their transmittance instead of $\tau_{atm}$. The transmittance $\tau_{rest}$ of water vapor+$CO_2$, which effectively occur below the ozone layer, is calculated using MODTRAN by setting the atmospheric ozone concentration to zero. The transmittance of ozone is calculated as follows:

$$\tau_{Ozone}(\theta,\lambda) = \tau_{atm}(\theta,\lambda) / \tau_{rest}(\theta,\lambda) \qquad (6)$$

As mentioned earlier, for the transmittance-based cosine model, it is reasonable to assume that the effective temperature of water vapor+$CO_2$ where it is highly absorptive is the ambient temperature $T_{amb}$, and the effective temperature of the completely masked ozone layer beyond is 0 K. It thus remains to calculate the effective temperature $T_{Ozone}$ of the ozone layer in the LWIR, and $T_{rest}$ of water vapor and $CO_2$ in the LWIR and in the 16-20 µm range. To do so, we first obtain from MODTRAN the effective hemispherical irradiance of $CO_2$ and water vapor $I_{rest}$ (i.e. without ozone), and that of the ozone layer, $I_{Ozone}$ by subtracting $I_{rest}$ from the hemispherical irradiance of the whole atmosphere ($I_{atm}$). In parallel, we also calculate the hemispherical emittance of water vapor+$CO_2$, and the ozone layer from the directional values $\varepsilon_{rest}(\lambda,\theta)$ and $\varepsilon_{Ozone}(\lambda,\theta)$ calculated earlier. We then solve for $T_i(\lambda)$ using the equation:

$$I_i(\lambda) = \varepsilon_i(\lambda) \int_0^{2\pi} \int_0^{\frac{\pi}{2}} 2hc^2\lambda^{-5} \left(e^{\frac{hc}{\lambda k_B T_i(\lambda)}} - 1\right)^{-1} \cos\theta \sin\theta \, d\theta d\varphi \qquad (7)$$

Plots of $T_{rest}$ and $T_{Ozone}$ are presented in Fig. 2 for the six MODTRAN standard atmospheres, US standard, Tropical, Midlatitude summer, Midlatitude winter, Subarctic summer and Subarctic winter in Fig. 2. As shown, the effective temperatures of the gaseous components are drastically lower than $T_{amb}$, with ΔT(lambda) being 50-100°C for ozone and 5-20°C for water vapor+$CO_2$, which the traditional transmittance-based model does not capture.

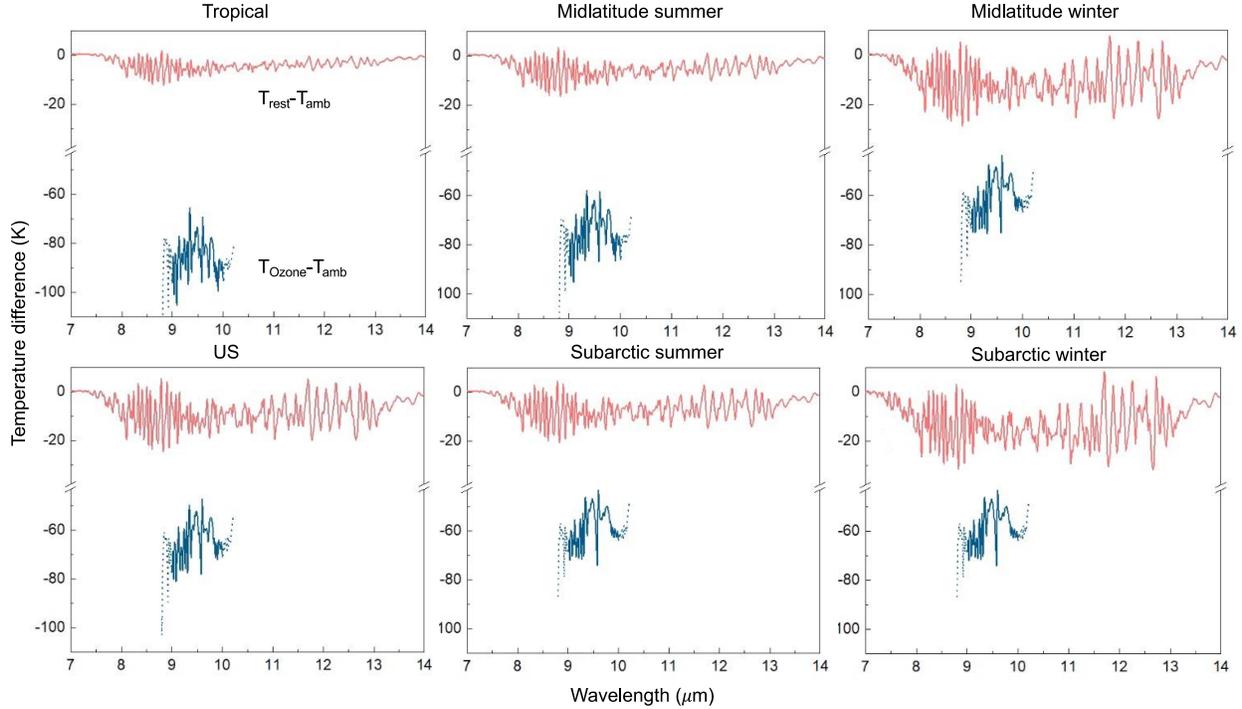

**Figure 2.** $\Delta T_{ozone}$ and $\Delta T_{rest}$ for the six MODTRAN standard atmospheres: Tropical, Midlatitude summer, Midlatitude winter, US 1976 Standard Atmosphere, Subarctic summer, Subarctic winter.

Equation 5, along with Equations 2-4, yields the directional irradiance, which in turn can be used to calculate the hemispherical irradiance ($I_{atm}$) that corrects the overestimation of the transmittance-based model. As an illustration, we present the resulting hemispherical sky irradiances $I_{atm}$ for the six MODTRAN standard atmospheres against the respective MODTRAN and transmission-model outputs (Fig. 3). As expected, the corrected transmission-based $I_{atm}$ is far closer to the MODTRAN irradiance than the original transmission-based approximation. More importantly, our method maintains its closeness to the MODTRAN model when the temperature $T_{amb}$ is changed. Fig. 4 shows the cooling potential $P_{cooling} = I_{BB}(T_{amb}) - I_{atm}$ of the traditional transmission-based model, the corrected model we propose, and the MODTRAN irradiances of versions of the standard atmospheres at different temperatures. The values of $I_{atm}$ are calculated using $T_{amb}$ for the traditional transmission-based model, using $T_{amb}$ and ΔT in Fig. 2 (assuming that ΔT is unaffected by $T_{amb}$) for the rectified model, and by scaling the irradiances of the standard atmosphere by the blackbody spectra corresponding to $T_{amb}$ for the MODTRAN model.

As shown in Fig. 4, the relative to the MODTRAN model, the traditional approximation underestimates the cooling potential of radiative coolers by 6 to 24 Wm$^{-2}$ or 12 to 29 % depending on the temperature and atmosphere type. The underestimations are large, particularly for high values of $I_{atm}$. Our corrected model, by comparison, is within 1-8 Wm$^{-2}$ or 0.1-7 % of the MODTRAN model, irrespective of the temperature, and the different total precipitable water and other greenhouse gas levels represented by the different atmospheres. The corrected transmission-based model thus provides an accurate irradiance relative to MODTRAN, while also providing angle-resolved irradiance values that the traditional transmission-based model provides.

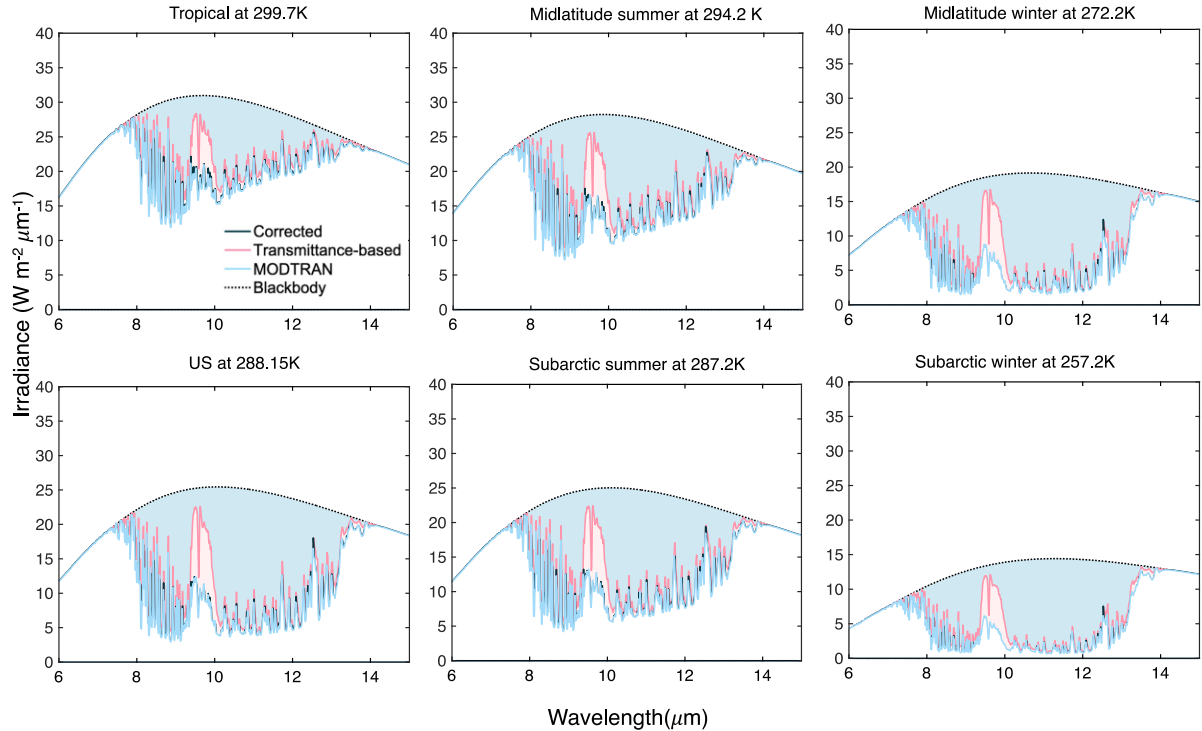

**Figure 3**. Spectral hemispherical atmospheric irradiance $I_{atm}$ for the transmittance-based model, our corrected model, the MODTRAN model, and the blackbody irradiance at $T_{amb}$ for the six different climates: Tropical, Midlatitude summer, Midlatitude winter, US 1976 Standard Atmosphere, Subarctic summer, Subarctic winter

### Discussions

Two results stand out from our analysis. From a computational standpoint, our results indicate that a higher cooling power can be achieved by radiative coolers than is calculated using the traditional transmittance-based model. This adds to their promise for cooling at local and global scales, with ~20% greater cooling potentials possible for some atmospheres. However, our results also suggest that except in rare cases, experimentally demonstrated radiative coolers perform less well relative to what is theoretically possible than previously thought. One reason for this could be that such radiative coolers often have their near-normal emittances reported and used in radiative cooling calculations, rather than their true hemispherical emittance, which may require full angularly-resolved measurements. Since hemispherical emittances are usually considerably lower than the near-normal emittances typically reported,[33] this could explain why this mismatch has not been previously noted.

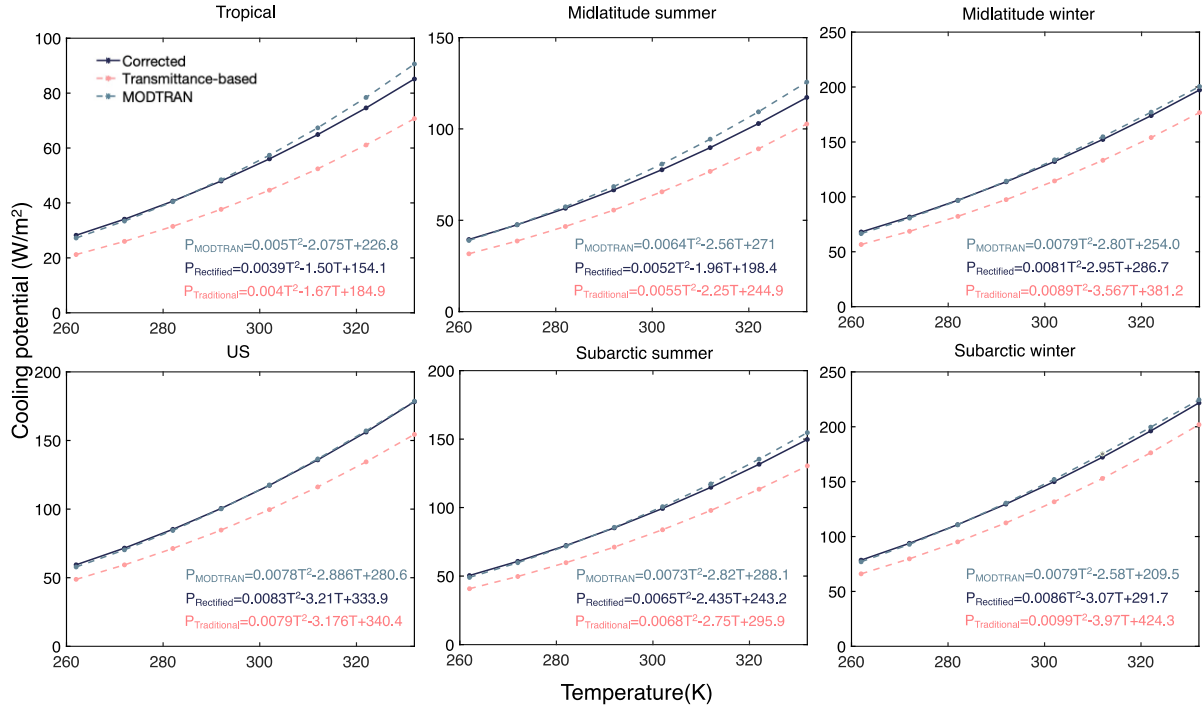

**Figure 4**. Maximum cooling potential ( $P_{cooling} = I_{BB}(T_{amb}) - I_{atm}$ ) calculated using the traditional transmittance-based model, the corrected model and MODTRAN for the six standard atmospheric profiles, but with $T_{amb}$ varied from 262 to 332 K. Dots indicate the calculated data, and the lines indicate fits with $R^2$>0.99. Equations for the fits are provided. The data at low temperatures for the Tropical, Midlatitude Summer and Subarctic summer atmospheric profiles should be used with caution because at such temperatures, the atmosphere does not typically hold the high TPW levels that characterize the standard cases.

**Table 1**. Analytical expressions of the $P_{cooling}$ corrections between the MODTRAN, transmittance-based and corrected models.

| $\Delta P_{cooling}$ | Tropical | Midlatitude Summer | Midlatitude Winter | US 1976 Standard | Sub-Arctic Summer | Sub-Arctic Winter |
|---|---|---|---|---|---|---|
| $P_{MODTRAN}$-$P_{Corrected}$ | $0.0011T^2$-$0.57T$+72.7 | $0.0012T^2$-$0.6T$+72.6 | $-0.002T^2$+$0.15T$-32.7 | $-0.0005T^2$+$0.32T$-53.32 | $0.0008T^2$-$0.385T$+45 | $-0.0007T^2$+$0.49T$-82.2 |
| $P_{MODTRAN}$-$P_{Traditional}$ | $0.001T^2$-$0.4T$+41.9 | $0.0009T^2$-$0.31T$+26.1 | $-0.001T^2$+$0.767T$-127.2 | $-0.0001T^2$+$0.29T$-59.8 | $0.0005T^2$-$0.071T$-7.9 | $-0.002T^2$+$1.39T$-214.8 |

Given the above implications of our work, we believe that it may be useful to contextualize prior works which used the transmittance-based model, and that future works should account for the underestimation of the theoretical radiative cooling potential by the traditional transmittance-based model. Towards that end, we have provided analytical expressions of the cooling potentials for the different MODTRAN atmospheres as a function of temperature, as well as analytical expressions of $\Delta P_{cooling}$ between the MODTRAN, transmittance-based and corrected models, in Table 1. Additionally, we have also made numerical data for the zenith-ward transmittances and $\Delta T_i$ for different model atmospheres, which can be used with our method to calculate angular and hemispherical emittances, publicly available.[34,35] We hope that these resources will be useful to researchers modelling atmospheric irradiances for radiative cooling applications.

## Author Contributions

J. M. conceived and determined the mathematical methodology for this work. X. H. and J. M. performed the calculations. A.P.R. supervised the study. J.M., X.H., and A.P.R. wrote the manuscript.

## Acknowledgements

J. M. was supported by Schmidt Science Fellows, in partnership with the Rhodes trust. We also acknowledge support of the UCLA Hellman Fellows Award.